# A novel spacetime concept for describing electronic motion within a helium atom

Kunming Xu
Environmental Science Research Center
Xiamen University, Xiamen 361005
Fujian Province, PR China

**Abstract**: Euclidean space and linear algebra do not characterize dynamic electronic orbitals satisfactorily for even the motion of both electrons in an inert helium atom cannot be defined in reasonable details. Here the author puts forward a novel two-dimensional spacetime model from scratch in the context of defining both electrons in a helium atom. Space and time are treated as two orthogonal, symmetric and complementary quantities under the atomic spacetime. Electronic motion observed the rule of differential and integral operations that were implemented by dynamic trigonometric functions. It is demonstrated that the atomic spacetime is not a linear vector space with Newtonian time, and within which calculus has non-classical definition, and complex wave functions have fresh physical significances. This alternative approach is original, informative and refreshing but still compatible with quantum mechanics in the formulation. The description of electronic resonance in helium is also comparable with classical mechanics such as an oscillating pendulum and with classical electromagnetism such as an LC oscillator. The study has effectively unified complex function, calculus, and trigonometry in mathematics, and provided a prospect for unifying particle physics with classical physics on the novel spacetime platform.

## 1. Introduction

Since antiquity, human creatures have believed that they know about space and time because of their direct experience, but scientific conception of space and time turns out to be elusive. While their meanings seem intuitively clear, attempts to define them encounter remarkable difficulty. This is not because space and time are so complex but they are so fundamental that there is not any preceding rule for reference. Like the stage of an act, the theme of a song, or the context of a paragraph, space and time are the foundation and background of all sciences and their importance can never be overestimated. How we define space and time actually determines our perspective and standpoint, from which we gain our worldview of the surrounding environment. It goes without saying that any heavy-duty successful theories must cope with the basic concepts properly. Significant examples are Newtonian mechanics and Einstein's relativity.

The ancient Egyptians learnt to measure lands and constructed pyramids thousands of years ago. By Newton's time, people had been pretty good at elementary Euclidean geometry But what was about time? Can we model time using some version of Cartesian coordinate system? Isaac Newton thought about this and defined that "Absolute, true and mathematical time, of itself, and from its own nature, flows equably, without relation to anything external." Given such a postulate, he developed three laws of motion that form the core of classical mechanics. Simple as it was, yet it has been the most practical and fruitful interpretation of space and time since then.

Space and time were customarily considered to be separate quantities until the 1900s when Minkowski proposed a four-axis spacetime continuum, in which a time dimension is coupled together with three space dimensions through events. Minkowski's four-axis coordinates became the framework for Einstein's special theory of relativity, which explains that space is relative, and time is relative, too. They are relative in the sense that when an object travels at a high speed, its time dilates and length contracts compared with those of rest objects. Such an unusual conclusion is demonstrated by Lorentz's transformations connecting an inertial reference frame and another frame moving at a constant velocity relative to it. In his general theory of relativity, Albert Einstein further proposed that time curves in the space around stars and other massive objects and came up with an equation relating the curvature tensor of the distance function to the distribution of matter and energy in spacetime.

Despite all progress, modern conception of spacetime originated from and is still limited by intuitive observations that are framed by Euclidean geometry and Newtonian time. Euclidean geometry does not incorporate time axis into the three-dimensional space framework, hence time remains isolated from space. Time is assumed to be one-dimensional and hence is not considered to be the counterpart of or in symmetry to space that I have proposed previously [1, 2]. Moreover, the three-dimensional space by Cartesian coordinates is a mathematical abstraction dissociated from real objects; and the recording of time by the ticking of a mechanical clock is an idealized counter dissociated from living organisms, the natural subjects of time sensing and recording. Such detachment and ideal abstractions of space and time must have limited their applicability in reality.

Classical mechanics works perfectly on the ground, but it breaks down when applied to a large astronomical distance. For example, objects are no longer traveling in a straight line as predicted by Newton's first law of motion, instead geodesics represent the paths of freely falling particles in a given cosmic space. Relativity takes over in explaining the discrepancy. Thus the extension of classical mechanics to curved space is invalid. Moreover, turning our focus to the microscopic world, the principle of uncertainty precludes the application of Newton's laws either. Quantum mechanics has to be called for to characterize the behavior of electrons. Oddly, quantum mechanics adopts statistical probability to describe electronic orbitals, and this approach is radically different from that of classical mechanics. And unfortunately, relativity and quantum physics still remain detached and to be unified. Since space and time are the underlying quantities of all physical phenomena, the limitation of classical mechanics and the wanting unification of relativity and quantum mechanics prompt us to search for new alternative space and time concepts on a more profound level and in a wider scope.

## 2. Redefining two dimensions by complex functions

If we admit that somewhere in the particle microcosm or in the universe, space and time might be different from what we are used to, then we need to be cautious on any presumptions that we have inadvertently introduced since civilizations. Let's discard every antecedent belief and premise except saying that space and time are a pair of fundamental physical quantities. To expatiate on what they are, we notice that there are two electrons within a helium atom. Hence we may identify space with one electron, and time with the other, or in a more general manner as will be introduced. Helium shell is such a conservative system that both electrons should best represent the two basic dimensions. It is by this approach that we explore the property of space and time through the description of the atomic structure as follows.

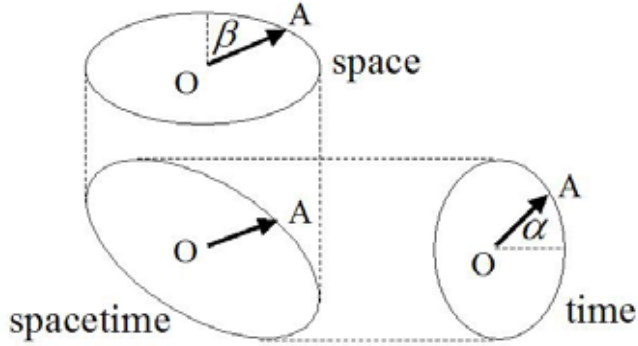

Figure 1. Oscillation movement of an electron within a helium atom where its time factor was determined by the rotation of α angle and space factor by the rotation of β angle.

Let's consider the motion of electrons in a helium atom. An electron had space and time properties, or say that an electron lived in a two-dimensional spacetime. As shown in Figure 1, an electron was confined to a space sphere as well as to a time sphere, both inseparable in that they were inherently coupled together forming the electron. Space and time cannot be divorced from the electron in helium shell as if a balloon will shrink without the inflation of the air. If the magnitude of the shrinking force was proportional to the space or time quantity that the electron possessed, then the motion of the electron was best described by the general harmonic oscillation. Such an oscillation without damping could be expressed by

$$\frac{d^2\Psi}{dt^2} = -\omega^2 \Psi, \qquad (1)$$

where $\Psi$ was time component of the electron and $\omega$ denoted angular velocity. One is tempted to compare this equation with Schrödinger's equation, which is much more complicated. But bear in mind that we were dealing with two-dimensional spacetime where space and time had non-classical meanings, so did their related wave functions and parameters. We shall discuss Schrödinger's equation in section 5 and suffice it to say here that the complexity of Schrödinger's equation beyond equation (1) is unnecessary in the context of the two-dimensional spacetime. Equation (1) provides much useful information on the electronic orbital with its typical solution as

$$\Psi = C_1(\cos\alpha - i\sin\alpha), \qquad (2)$$

where $C_1$ was a time constant and $\alpha$ was a radian angle related to time component of the electron, which satisfies the relation of

$$-\frac{d\alpha}{dt} = \omega, \qquad (3)$$

where the negative sign implied the contrary aligning direction of dynamic $\alpha$ angle movement relative to the time dimension orientation. The solution can be verified by carrying out differential operations twice on $\Psi$ with respect to time $t$.

Similarly, since space and time were relative and symmetric, the wave function describing space component of the electron was:

$$\frac{d^2\psi}{dl^2} = -\frac{1}{r^2}\psi, \tag{4}$$

$$\psi = C_2(\cos\beta + j\sin\beta), \tag{5}$$

$$\frac{1}{r} = \frac{d\beta}{dl}, \tag{6}$$

$$v = \omega r, \tag{7}$$

where $C_2$ was a space constant, $j$ was a complex number notation like $i$ but it described imaginary space instead of imaginary time component, $r$ was the orbital radius, and $v$ was velocity.

Because an electron must possess time and space components altogether, and both components were orthogonal at any moment, we expressed the electronic wave function by the product of both components as

$$\Omega = \Psi\psi, \tag{8}$$

whence

$$\Omega = C_1C_2(\cos\alpha\cos\beta - i\sin\alpha\cos\beta - ij\sin\alpha\sin\beta + j\cos\alpha\sin\beta). \tag{9}$$

Here multiplication operation made sense for the coordination of two orthogonal quantities. However, to understand the physical meaning of the wave functions, we must decipher the meanings of $i$ and $j$ notations. In principle, a complex number is introduced when real numbers cannot express a conventional two-dimensional vector. Under Cartesian X-Y coordinates, the identifier $i$ is an operator casting a real number in X-axis into an imaginary component along Y-axis. Logically, the imaginary and the real parts of a complex number belong to different space orientations when representing a vector.

How to physically express various spacetime dimensions in two-dimensional helium shell? If we started with a dimensionless quantity, changed it in the direction of reducing a time dimension and increasing a space dimension, and reversed the other way around to complete a cycle, then we got four types of dimensional quantities out of the possible one-dimensional space and one-dimensional time combinations as represented by (1, $\omega$, $v$, $r$) with SI units of (1, 1/s, m/s, m) respectively, adopting meter and second to denote space and time units. Thus, the significance of complex notations in equation (9) was interpreted physically as

$$\begin{pmatrix} 1 & i \\ ij & j \end{pmatrix} = \begin{pmatrix} 1 & \omega \\ v & r \end{pmatrix}. \tag{10}$$

Because $\Psi$ and $\psi$ represented two orthogonal space and time components, a partial differentiation on $\Omega$ with respect to space or time dimension only affected its space or time factor, e.g.

$$\frac{\partial^2 \Omega}{\partial t^2} = \psi \frac{d^2 \Psi}{dt^2},  \quad (11)$$

$$\frac{\partial^2 \Omega}{\partial l^2} = \Psi \frac{d^2 \psi}{dl^2}.  \quad (12)$$

Combining equations (1), (4), (7), (11), and (12) yields a two-dimensional oscillation equation for electrons within helium shell:

$$\frac{\partial^2 \Omega}{\partial t^2} = v^2 \frac{\partial^2 \Omega}{\partial l^2},  \quad (13)$$

which we shall call duality equation. Upon separation of space and time variables, this partial differential equation would revert to two ordinary differential equations (1) and (4). A linear differential equation like this has a very important property that if $\Omega_0$ and $\Omega_1$ are two valid solutions to it, then $a_0 \Omega_0 + a_1 \Omega_1$ is also a valid solution where $a_0$ and $a_1$ are constant parameters. Because there are four terms on the right-hand side of equation (9), we shall regard every term, $\Omega_i$, as a characteristic root to equation (13) without losing any dimensions physically:

$$\begin{pmatrix} \Omega_0 \\ \Omega_1 \\ \Omega_2 \\ \Omega_3 \end{pmatrix} = C_1 C_2 \begin{pmatrix} \cos\alpha \cos\beta \\ -\omega \sin\alpha \cos\beta \\ -v \sin\alpha \sin\beta \\ r \cos\alpha \sin\beta \end{pmatrix}. \quad (14)$$

However, there are only two electrons within a helium atom, so each electronic orbital must include two adjacent roots. For example, an electronic wave function within helium shell may take the form of ($\Omega_0 + \Omega_1$) while the other ($\Omega_2 + \Omega_3$). Thus we have defined both electrons in helium shell by two complex functions from the general harmonic oscillation equation.

**3. Duality in helium**

Between the four roots of duality equation, there were strict differential or integral relationships:

$$-\frac{\partial \Omega_0}{\partial t} = \Omega_1, \quad (15)$$

$$\int \Omega_1 dl = \Omega_2, \quad (16)$$

$$-\int \Omega_2 dt = \Omega_3, \quad (17)$$

$$\frac{\partial \Omega_3}{\partial l} = \Omega_0. \tag{18}$$

How to interpret these relationships? We believed that each of these mathematical equations had its corresponding physical process that reflected electronic behavior in dynamic action. To specify, when an electron was in the state of ($\Omega_0 + \Omega_1$), its $\Omega_0$ component was transforming into $\Omega_1$ component according to equation (15). As this process completed, the integral process of equation (16) started off. Or it might well be that the processes of equations (15) and (16) were undergoing simultaneously. In short, the electron was shifting its state from ($\Omega_0 + \Omega_1$) to ($\Omega_1 + \Omega_2$) losing a time dimension and gaining a space dimension while the other electron evolved from state ($\Omega_2 + \Omega_3$) to state ($\Omega_3 + \Omega_0$) increasing a time dimension and reducing a space dimension. Hence the electrons in helium shell were not static, but were switching states continuously and cyclically.

By the simplest interpretation, the oscillation of electrons was somewhat similar to a pendulum where a body suspended from a fixed support swings freely back and forth under the influence of gravity. One significant difference was that the period of the electronic cycle was very short. If each electron were orbiting around the nucleus like planets around the sun physically as was suggested by Niels Bohr, then they would emit energy due to their high frequency. As a result, the system could not maintain conservative, and the oscillations would be damped quickly. To overcome this, the electrons must oscillate through changing states so that both electrons exchanged energy internally, i.e. each electron received the momentum and energy emitted by the other. Thus the system would not lose energy to the outer environment so that the oscillatory cycles proceeded forever. Here the electron was revolving in the sense that it changed physical state continuously and periodically as the state point, *A*, orbited around the origin O (Figure 1). The circular track of point *A* represented the pathway of electronic state transformation or induction rather than kinematic movement.

Combining equations (15) and (16) together, and (17) and (18) together yields:

$$-\frac{\partial \Omega_0}{\partial t} = \frac{\partial \Omega_2}{\partial l}, \tag{19}$$

$$-\int \Omega_0 dl = \int \Omega_2 dt, \tag{20}$$

which meant that the changing rate of one electron in time was compensated by the varying rate of the other electron in space. The derivative form conforms to Faraday's law,

$$\nabla \times E = -\frac{\partial B}{\partial t}. \tag{21}$$

Since there were only a space and a time dimensions in the duality shell, the *curl* operator, $\nabla \times$, declined to $\partial/\partial l$. If we treated $\Omega_0$ as a magnetic field and $\Omega_2$ as an electric field, and recognized that space and time were varying in opposite directions, then equation (19) was indeed the expression of Faraday's law. This indicated that electronic oscillation was an electromagnetic phenomenon as electrons were exchanging energy. Because $\Omega_0$ and $\Omega_2$ were not necessarily be the magnetic field and the electric field strengths, we might say in a more general term that the electrons were in certain states obeying a duality principle similar to Faraday's law, one of Maxwell's equations. For example, if we treated $\Omega_0$ as a probability density function and $\Omega_2$ as probability current, then equation (19) also indicated that a change in probability density in region *l* was compensated by a net change in flux into that region. This also agreed with quantum mechanics on probability.

Considering $\omega^2 = -1$ and $r^2 = -1$ dimensionally from equation (10), we may rewrite the calculus relationships of (17) and (18) as:

$$-\frac{\partial \Omega_2}{\partial t} = \Omega_3, \tag{22}$$

$$\int \Omega_3 dl = \Omega_0. \tag{23}$$

The reason for their equivalence was that in the two-dimensional world, reducing a time dimension from a wave function for the second times was equivalent to increasing it while increasing a space dimension from a wave function for the second times was equivalent to reducing it. This is somewhat analogous to the situation of a man who walks along a circle. As he walks forward a distance of half circumference, he arrives at the other end of the diameter; but as he continues to go forward for another distance of half circumference, he returns to his original place, forming a loop. In the same manner, equations (15), (16), (22), and (23) formed a close loop where reducing a time dimension alternated with increasing a space dimension, e.g. the process of increasing a space dimension was always accompanied by the process of decreasing a time dimension, or vice versa. In other words, expansion of space was undergoing with release of time wrinkles whereas contraction of space resulted in condensation of time. When space fully unfolded, it lost all density and wrapped back spontaneously according to the stipulated cycle, so did time. Space and time components were coupled together in such an intimate way that spacetime was a finite and yet unbounded continuum. This spacetime view agreed well with the theory of relativity that spells out time dilation and length contraction.

From the perspective of waves, if we regarded each root, $\Omega_i$, as a waveform, then each electronic orbital was composed of two adjacent roots, and therefore had two waveforms. Because every pair of adjacent roots were exactly one dimension apart, separated by either a time or a space dimension, the two waveforms were orthogonal to each other. In other words, an electron manifested itself as two perpendicular waveforms. For instance, an electron may exist as a pair of interwoven electric wave and magnetic wave. An electronic wave propagated from one waveform to another following the differential and integral rule. How to explain electronic duality? We believed electrons were real particles, but they may not exist as solid particles in atoms all the time. No one has ever captured a single electron in its static particle form as biologists often capture a bacterium under the microscope. If we associated time with a condensed stable particle and space with an expanded volatile cloud medium, then the electronic oscillation can be interpreted as being transforming between solid particle and electron cloud. Electrons exhibited wave and particle behavior because of their varying states between space and time.

**4. Symmetry, orthogonality, and calculus in the atomic spacetime**

We shall encounter many unconventional consequences mathematically in the atomic spacetime. Firstly, there were two main aspects when we said that space and time were symmetric previously. On the one hand, space and time symmetry indicated that each was the counterpart of the other, i.e. they were coordinative and hence permitted multiplication between their components as was done in equation (8). In the two-dimensional spacetime, the coordinative property of both space and time dimensions also expressed their complementarity. On the other hand, from equations (19) and (20), we considered $\Omega_0$ and $\Omega_2$ to be symmetric regarding differential operations with respect to a time and to a space dimension as well as regarding integral operations over a space and over a time dimension. Symmetry meant the calculus relationships for electronic transformation along $t$ and $l$ dimensions. Thus symmetry had new physical significances in the atomic spacetime.

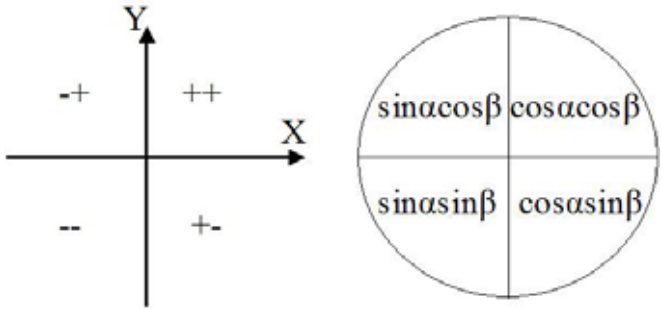

Figure 2 Sine and sign in four contiguous quadrants

Secondly, it is interesting to compare sine/cosine functions of the four terms in equation (14) with -/+ signs in four Cartesian quadrants (Figure 2), but instead of saying that $\Omega_0$ and $\Omega_1$ were in symmetry, we regarded them as orthogonal because they were exactly one dimension apart as was shown in equation (15). Orthogonality was defined as exact differential and/or integral relationship between two wave functions. By this definition, we deduced that the four characteristic roots of duality equation were all mutually orthogonal from equations (15) to (18). In the atomic spacetime, orthogonal quantities were perpendicular with a π/2 radian phase difference between their trigonometric functions, but they were transforming from one quantity to the other according to calculus rule. This was radically different from orthogonality concept in Euclidean space where orthogonal quantities are supposed to be mutually independent and inexchangeable.

Thirdly, since electronic motion followed differential and integral operations transforming from one state to another, the plus sign in the complex function $(\Omega_0 + \Omega_1)$ was mathematically synonymous with the differential operation of equation (15). By this definition of addition operation, wave function ($\Omega_0 + \Omega_1$) had different physical meaning from ($\Omega_1 + \Omega_0$) so that wave function addition did not observe the commutative law and hence the atomic space was not a linear vector space. The atomic spacetime was governed by the rule of differentiation and integration instead of minus and plus operations. While multiplication denoted coordination between symmetric quantities, addition indicated calculus transformation between orthogonal quantities.

Because in the atomic spacetime, differential and integral operations were the rule for electronic transformation, we need to examine the significance of calculus in the dynamic process. A differential operation upon a trigonometric function with respect to a time dimension, such as that dictated by equation (15), did not physically happen in a flash, but it was carried out gradually and smoothly. For example, when $\cos\alpha$ received the differentiation command, the angle $\alpha$ was then rotating gradually up to ($\pi/2 + \alpha$), at which point the differential operation completed so that $\cos\alpha$ transformed into $-\sin\alpha$ accordingly. In the atomic spacetime, we defined the calculus of trigonometric function as follows:

$$\frac{d}{d\alpha}\cos\alpha = \cos(\alpha + \frac{\pi}{2}) ; \tag{24}$$

$$\frac{d}{d\alpha}\sin\alpha = \sin(\alpha + \frac{\pi}{2}) ; \tag{25}$$

$$\int \cos\beta \, d\beta = \cos(\beta - \frac{\pi}{2}) ; \tag{26}$$

$$\int \sin\beta \, d\beta = \sin(\beta - \frac{\pi}{2}) . \tag{27}$$

The correctness of these expressions can be easily verified under conventional trigonometric calculus. But we here interpreted the plus and minus signs in these equations as a dynamic and continuous angle increasing or decreasing up to π/2 radian displacement. This dynamic definition was remarkably different from infinitesimal calculus. Under the context of electronic wave functions, we implemented equations (15) and (16) by:

$$-\frac{d}{dt}\cos\alpha = -\omega\cos(\alpha + \frac{\pi}{2}) ; \tag{28}$$

$$\int \cos\beta \, dl = r\sin(\beta - \frac{\pi}{2}) . \tag{29}$$

As shown in Figure 3(a), the transformation of the time component of $\Omega_0$ from $C_1\cos\alpha$ to $-C_1\omega\sin\alpha$ can be expressed as the dynamic motion of point C along semicircular arc ACB. At any specific point C, chord BC denoted $C_1\cos\alpha$ while chord AC denoted $-C_1\omega\sin\alpha$ where factor $\omega$ was a complex number notation indicating mutually perpendicular relationship between BC and AC. As radian angle $\alpha$ rotated from 0 to π/2, chord BC disappeared while chord AC increased to the maximum of diameter AB. This was a geometric interpretation of differential operation $-\partial\Omega_0/\partial t$ where the minus sign indicated time component decreasing. It goes without saying that the velocity of $\alpha$ rotation, or angular velocity $\omega$, determined the speed of the differential process and hence the period of electronic oscillatory cycle.

Likewise, Figure 3(b) illustrated the geometric course of equation (29). As radian angle $\beta$ decreased from π/2 to 0, point C tracked from A to B along the semicircular arc ACB. At any specific point C along the pathway, chord AC denoted $C_2\cos\beta$ while chord BC denoted $C_2 r\sin\beta$, both being perpendicular at any time as noted by factor $r$ as complex number identifier. Orthogonality meant AC⊥BC at any moment even though they were transforming between each other dynamically. Thus integral operation of equation (16) was implemented by continuous $\beta$ angle decreasing in the wave function. When $\alpha$ and $\beta$ angle were in complement and synchronized at any time, equation (19) held describing the relationship between $\Omega_0$ and $\Omega_2$ wave functions. In this way, we have explained electronic motion by calculus and the implementation of calculus by trigonometry through continuous radian angle rotation.

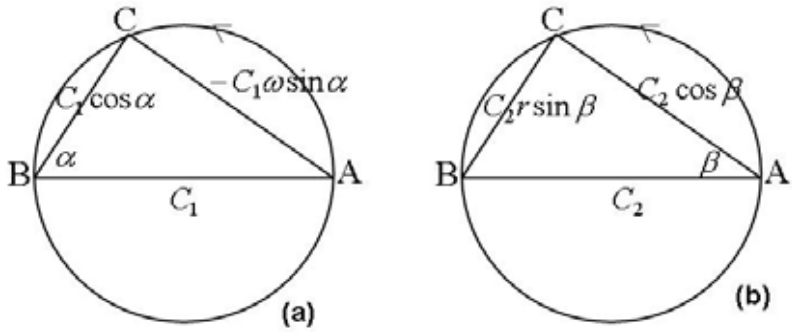

Figure 3. Trigonometric interpretation of dynamic calculus for the motion of electrons with concurrently changing (a) time component and (b) space component.

Mathematical expressions represented dynamic physical processes. In light of the continuous evolution of electronic wave functions in space and time, two electrons at a specific moment must constitute two basic dimensions, time and space, which were separated by a phase of π/2 in their waveforms. For example, an electron ($\Omega_0+\Omega_1$) may represent time as $\alpha$ equals 0 while the other electron ($\Omega_2+\Omega_3$) may represent space as $\alpha$ equals π/2. At that specific moment, both $\Omega_1$ and $\Omega_3$ vanished so that one electron, $\Omega_0=C_1$, indicated a full time dimension while the other, $\Omega_2=C_2$, indicated a full space dimension. When 0<$\alpha$<π/2, both electrons contained a mix of space and time components according to the specification. Since both electrons in helium shell at a certain phase represented space and time dimensions and they were symmetric and orthogonal at any time, we concluded that space and time dimensions were both symmetric and orthogonal as well.

**5. Duality equation versus Schrödinger's equation**

Our description of electronic orbitals was not only self-consistent but also compatible with well-established quantum physics. This section integrates duality equation with Schrödinger's equations. We shall derive a common heat equation from both two-dimensional spacetime concept in helium shell and Schrödinger's one-dimensional equation.

In the atomic spacetime, because space and time components were relative and symmetric, the derivative of a wave function $\Omega$ with respect to time and its derivative with respect to space must be equal:

$$\frac{\partial \Omega}{\partial t} = v \frac{\partial \Omega}{\partial l}, \tag{30}$$

where $v$ compensated the dimensional difference. Furthermore, since two electrons within a helium atom were converting between each other, the quantity of one electron was proportional to the rates of changes in another electron and in itself in the dynamic flow. Let $\Omega_0$ and $\Omega_2$ be the two electrons. We had a quantitative relation concerning dynamic budget equilibrium:

$$\Omega_0 = a \frac{\partial \Omega_0}{\partial t} - b \frac{\partial \Omega_2}{\partial t} + c, \tag{31}$$

where $a$, $b$, and $c$ were constant parameters. Since space and time were symmetric, when one electron wholly occupied space $\Omega_0$, another electron $\Omega_2$ should be exactly full time component as was mentioned in the previous section. Appling this special boundary condition to equation (31), we got zero for the first and the third terms so that we had

$$\Omega_0 = -b \frac{\partial \Omega_2}{\partial t} \tag{32}$$

in its simplest form. Since at this moment, space and time, as represented by $\Omega_0$ and $\Omega_2$, respectively, were symmetry and must have a similar shape in mathematical expression, we therefore rewrited equation (32) as:

$$\Omega = -\frac{1}{\omega} \frac{\partial \Omega}{\partial t}, \tag{33}$$

where $\omega$ was a dimension compensator denoting a reciprocal time dimension. Comparing equations (30) and (33), we also got

$$\Omega = -r \frac{\partial \Omega}{\partial l}. \tag{34}$$

Substituting $\Omega$ value of this equation into the right-hand side of equation (30) produces

$$-\frac{1}{\omega} \frac{\partial \Omega}{\partial t} = r^2 \frac{\partial^2 \Omega}{\partial l^2}. \tag{35}$$

This equation is in the shape of a well-known thermal diffusion equation or heat equation. It is also called Fick's second law when applied to characterize concentration or molecular diffusion, the diffusion coefficient being assigned to $-\omega r^2$ in this case. Thus it would be proper to say that electronic motion followed the diffusion law.

On quantum mechanics side, Schrödinger's equations for the motion of a particle are the starting point for the development of quantum theory. For a free electron, one-dimensional differential equation is as follows.

$$i\hbar \frac{\partial \phi}{\partial t} = -\frac{\hbar^2}{2m} \frac{\partial^2 \phi}{\partial x^2} . \tag{36}$$

In order to compare this equation with equation (35), we write down the following basic physical relationships:

$$r = \frac{\lambda}{2\pi}, \omega = 2\pi f , \tag{37}$$

$$E = hf , p = \frac{h}{\lambda}, \tag{38}$$

$$\hbar = \frac{h}{2\pi}, \tag{39}$$

$$E = \frac{p^2}{2m}, \tag{40}$$

where $h$, $\lambda$, $f$, $E$, and $p$ refer to Planck's constant, wavelength, frequency, energy, and momentum, respectively. For characterizing the kinematics of an oscillating object, parameters $\omega$ and $r$ are more descriptive and pertinent than $\hbar$ and $m$. They are closely related to energy and momentum through the rationalized Planck's constant:

$$\omega = \frac{E}{\hbar} ; \frac{1}{r} = \frac{p}{\hbar} . \tag{41}$$

After converting parameters $\hbar$ and $m$ into $\omega$ and $r$, equation (36) becomes

$$-\frac{1}{\omega} \frac{\partial \phi}{\partial t} = r^2 \frac{\partial^2 \phi}{i \partial x^2} . \tag{42}$$

Under the atomic spacetime, we interpreted the denominator $i\partial x^2$ as the generalized space dimensions $\partial l^2$ so that equations (35) and (42) were equivalent. The two space dimensions contained in $\partial l^2$ in equation (35) were orthogonal and had different meanings under Euclidean geometry. We may use $\partial x$ to indicate the first dimension of $\partial l$ within wave function $\phi$ and use $i\partial x$ to denote the successive space dimension $\partial l$ within $\phi$. In this sense, the complex number identifier represented the shifting of the space dimension order, and had the effect of rotating a space dimension to its perpendicular orientation under Euclidean geometry. This was in consistent with our original interpretation of complex number identifier $i$ or $j$ where it transformed its operand to the dimension orthogonal to it. Thus, the one-dimensional heat equation is the common ground of duality equation and Schrödinger's equations. After all, they diverge into different paths in equation formalism thereafter due to their different perceptions on spacetime.

Quantum mechanics extends one-dimensional Schrödinger's equation into three-dimensional by introducing Laplacian operator:

$$i\hbar \frac{\partial \phi}{\partial t} = -\frac{\hbar^2}{2m}\nabla^2 \phi, \qquad (43)$$

which sees three dimensions of space in X, Y, and Z orientations. By using Laplacian operator $\nabla^2$, it is implied that electronic orbitals distribute equally in X, Y, and Z directions, i.e., that space is homogenous and isotropic. Notwithstanding this impression, quantum mechanics traditionally handles Schrödinger's equation by transforming Cartesian coordinates into spherical polar coordinates pertinently and then try to derive quantum information under the constrains that the equation must have solutions and that the wave functions must be normalizable. Only via this transformation, it is successful in getting much useful information on electronic orbitals.

In contrast, since the second derivative contains the process of the first derivative, wave equation (35) is actually two-dimensional in helium spacetime. Moreover, because space and time were treated as symmetric in helium, we may further substitute the left-hand side of equation (35) by equation (33) so that time components have second derivatives too, which produces duality equation. Thus, duality equation agrees with Schrödinger's equation mathematically even though we have given it a fresh physical interpretation.

**6. Duality in an LC oscillator**

One will not be satisfied with the foregoing abstract description of electronic motion in the atomic spacetime. Neither quantum mechanics nor the standard model of particles and forces provides clearer explanation of electronic motion in the inert atom. We therefore investigate the property of an LC oscillator in this section with the hope that readers will gain better insight into the behavior of electrons in harmonic oscillation and the space and time dimensions that we have defined.

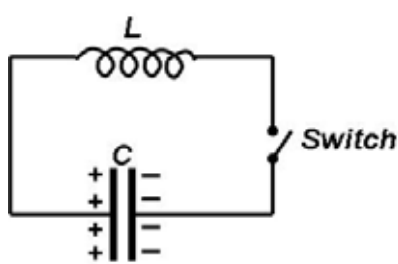

Figure 4. An *LC* circuit

Consider an idealized circuit like that shown in Figure 4 containing only a switch, a capacitor *C*, and an inductor *L* with *N* turns of a coil (and then treating *N*=1 for brevity), ignoring the resistance in the wire. Suppose the capacitor is initially charged so that one plate has positive charge *Q* and the other plate has negative charge of the same amount. As soon as the switch is closed, the capacitor discharges. The inductor initially opposes the growth of the current in the circuit by creating a change in magnetic flux through itself. The change in the magnetic flux induces an EMF in the circuit. After the capacitor discharges completely, the EMF drives a current in the circuit in the opposite direction, charging the capacitor in the reverse polarity. The cycle then repeats itself in the opposite direction. In the absence of any electric and magnetic energy loss the oscillations will continue back and forth indefinitely, which may be characterized by

$$\frac{d^2q}{dt^2} = -\frac{1}{LC}q, \tag{44}$$

where $q$ indicates charge in the capacitor varying in sinusoidal form:

$$q = Q\cos\omega t, \tag{45}$$

$$\omega = \sqrt{\frac{1}{LC}}. \tag{46}$$

In such an electromagnetic resonance, electric current $I$, EMF $V$, and magnetic flux $\Phi$ have the following basic relationships:

$$I = \frac{dq}{dt}, \tag{47}$$

$$V = -L\frac{dI}{dt}, \tag{48}$$

$$-N\frac{d\Phi}{dt} = V, \tag{49}$$

whence

$$I = -Q\omega\sin\omega t, \tag{50}$$

$$\Phi = -\frac{LQ\omega}{N}\sin\omega t, \tag{51}$$

$$V = LQ\omega^2\cos\omega t. \tag{52}$$

By the method of section 2, we may describe the harmonic oscillations in more physical dimensions than the function of electric charge only. The following two complex wave functions satisfy the oscillation equation in general:

$$q_1 = q + I$$
$$= Q(\cos\omega t - \omega\sin\omega t), \tag{53}$$

$$q_2 = \Phi + V$$
$$= QL\omega(-\sin\omega t + \omega\cos\omega t). \tag{54}$$

Here an electron might take the form of a static particle charge, flow as an electric current, transform into a magnetic flux, and build an electromotive force. We also realize that the charge is stored in the capacitor while the magnetic flux exists inside the inductor, both electronic components orthogonal in that a capacitor allows alternating current to pass but cuts off direct current whereas an inductor allows direct current to pass but impedes alternating current. Taking the hardware of the circuit into considerations, we may associate $q_1$ with $\cos\beta$ for a typical capacitor while associate $q_2$ with $\sin\beta$ for a typical inductor. As an electron undergoes electromagnetic oscillation, it takes a path from the capacitor, along the circuit, into the coiled wire of the inductor, through the inductor, back into circuit, and again to the capacitor. Besides the temporal sinusoidal waves, wave functions of $q_1$ and $q_2$ had spatial sinusoidal waves in terms of the shifting hardware association along the pathway. Since the temporal wave depended on the spatial transition along the circuit, both sinusoidal waves were concurrent and synchronized automatically. We adopted multiplication to express the hardware associations so that

$$q_1 = Q(\cos \omega t - \omega \sin \omega t)\cos \beta, \qquad (55)$$

$$q_2 = QL\omega(-\sin \omega t + \omega \cos \omega t)\sin \beta. \qquad (56)$$

Careful readers surely find that these two wave functions correspond to the two dimensions that we have defined previously by two electrons of $(\Omega_0 + \Omega_1)$ and $(\Omega_2 + \Omega_3)$. Here charge and current constituted a dynamic complex function associated with the capacitor while magnetic flux and EMF formed another associated with the inductor, both functions corresponding to two electrons in helium shell. We did not imply that both electrons in helium shell were in the same physical states as in the idealized LC oscillator, but it was certain that they were undergoing harmonic oscillations analogous to it more or less. Man can design an LC oscillator with trivial resistance in the circuit, but the idealized LC oscillator without damping as was described can only be created by nature, in the form of an inert atom. How wonderful the nature is! How much have we known about its secret?

**7. The atomic spacetime worldview**

Based on the mere assumption that electrons observed harmonic oscillation equation, we have given an orbital interpretation of space and time orthogonality in a two-dimensional world where time and space were treated as two symmetric and relative facets of electrons. Time and space were inherently associated with physical entities that instantiated them. Here space was no more a conventional three-dimensional volume, and time was no more a unidirectional flow. The significances of space and time were different from their usual meanings. When talking about these two dimensions, we were referring to two modes in spacetime as were instantiated by two electrons in their conservative system. It was more proper to adopt sine and cosine functions in multiple traditional dimensions (1, $\omega$, $v$, $r$) to characterize their intricacy than to use isotropic and linear X, Y, and Z coordinates independent of real objects. The atomic spacetime was continuous through differential and integral operations. This new outlook of spacetime was beyond our usual mental concept.

The atomic spacetime was calculus spacetime; the atomic spacetime was complex wave functions spacetime; and the atomic spacetime was trigonometric spacetime. We defined electronic motion by differential equations, expressed the wave functions by complex numbers; and implemented the dynamic processes by trigonometry in a coherent manner. But the atomic space was not a linear vector space, neither Euclidean space nor Hilbert space. As infinitesimal calculus and complex number drive mathematics towards idealism farther and farther, our new definition of dynamic calculus and physical interpretation of complex number under the the atomic spacetime brought it back to reality. Granting the new interpretation of spacetime did not mean to overthrow existing successful theories. Just as Einstein's relativity does not invalidate Newtonian mechanics, but sees objects in a wider scope and effectively extends it, our example here supplemented the description of electronic orbitals where classical physics ceases to be in effect and the power of quantum mechanics is so limited.

Because we viewed space and time in a helium atom differently from conventional three-dimensional space with one-dimensional time worldview, we were actually examining the world from a new perspective and by a new standard. Just as things may take various shapes from different angles, it was not surprising that the results we got might be quite different from those of quantum mechanics. The question is not which one is correct but which one is more informative and elegant to tackle the problem under consideration. Indeed, our theory could be established without challenging any successful theories, but adding a new insight to them instead.

## 8. References

[1] Xu, K. Discovering the spacetime towards grand unification, the theory of quaternity [M], Xiamen University Press, March 2005, pp.1-136.

[2] Xu, K. Novel Spacetime Concept and Dimension Curling up Mechanism in Neon Shell, http://xxx.lanl.gov/abs/physics/0511020.